\documentclass[aps,prl,twocolumn,groupedaddress,superscriptaddress]{revtex4}

\usepackage{graphicx}
\usepackage{colortbl}

\makeatletter
\def\btt#1{\texttt{\@backslashchar#1}}%
\DeclareRobustCommand\bblash{\btt{\@backslashchar}}%
\makeatother

\newcommand{\ETX}{$\theta$-(BEDT-TTF)$_2$X}

\newcommand{\aETI}{$\alpha$-(BEDT-TTF)$_2$I$_3$}

\newcommand{\RbZn}{$\theta$-RbZn}
\newcommand{\CsZn}{$\theta$-CsZn}
\newcommand{\I}{$\theta$-I$_3$}
\newcommand{\aI}{$\alpha$-I$_3$}

\newcommand{\rhoout}{$\rho_\perp$}
\newcommand{\rhoin}{$\rho_\parallel$}
\newcommand{\ani}{$\rho_\perp$/$\rho_\parallel$}

\begin{document}


\title{Anomalous 2D-confined electronic transport in layered organic charge-glass systems}

\author{Takuro Sato}
\email{takuro.sato@riken.jp}
\affiliation{Department of Applied Physics, University of Tokyo, Tokyo 113-8656, Japan}

\author{Kazuya Miyagawa}
\affiliation{Department of Applied Physics, University of Tokyo, Tokyo 113-8656, Japan}

\author{Masafumi Tamura}
\affiliation{Department of Physics, Faculty of Science and Technology, Tokyo University of Science, Noda, Chiba 278-8510, Japan}

\author{Kazushi Kanoda}
\email{kanoda@ap.t.u-tokyo.ac.jp}
\affiliation{Department of Applied Physics, University of Tokyo, Tokyo 113-8656, Japan}

\date{\today}

\begin{abstract}

To get insight into the nature of the electronic fluid in the frustration-driven charge glasses, we investigate in-plane and out-of-plane charge transport for several quasi-triangular-lattice organic systems, {\ETX} [X=RbZn(SCN)$_4$, CsZn(SCN)$_4$ and I$_3$]. 
These compounds host a charge order, charge glass and Fermi liquid, depending on the strength of charge frustration. 
We find that the resistivity exhibits extremely two-dimensional (2D) anisotropy and contrasting temperature dependence between in the in-plane and out-of-plane directions in the charge glass phase, qualitatively distinguished from the charge order and metallic states. 
The experimental features indicate that the frustration-induced charge glass carries an anomalous 2D-confined electronic fluid with possible charge excitations other than conventional quasiparticles.
\end{abstract}


\maketitle

A glass state of electrons not originating in disorder, a charge-glass (CG) state, arises from frustration in Coulomb interactions on triangular-like lattices \cite{Anderson1956, Mahmoudian2015, Rademaker2018, Rademaker2016,Yoshimi2012} similarly to in frustrated spin systems \cite{Shimizu2003, Balents2010, Zhou2017}, and is conceptually distinguished from conventional electronic inhomogeneity as caused by randomly located dopants \cite{Dagotto2005, Zeljkovic2012, Bogdanovich2002}. 
The CG was demonstrated in layered organic conductors with quasi-triangular lattices \cite{Kagawa2013, Sato2014, Sato2014b, Sato2016, Sato2017} through observing a charge inhomogeneity \cite{Sato2017, Chiba2008}, an exponential slowing down of electron dynamics \cite{Kagawa2013, Sato2014}, and ergodicity breaking signified by the cooling-rate dependence and aging of resistivity \cite{Sato2014} as well as electronic crystallization from glass \cite{Sato2017, Sasaki2017}; all of them are among the hallmark properties of glasses \cite{Debenedetti2001, Angell1991, Ediger2000}.

\begin{figure}
\includegraphics[width=8.9cm,clip]{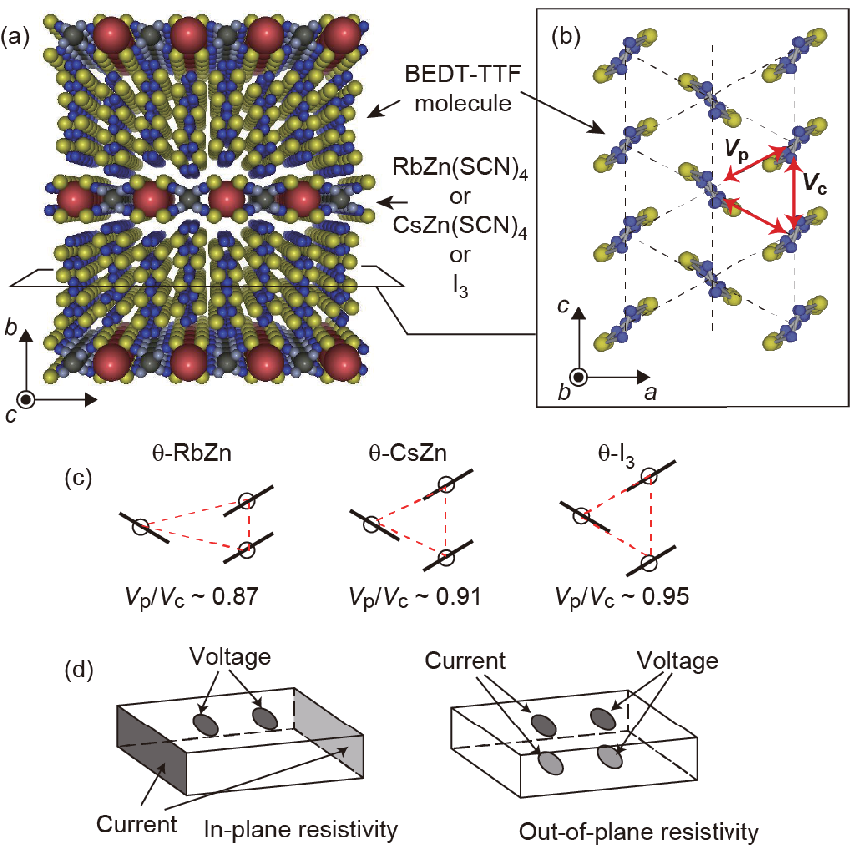}
\caption{(Color online) 
(a) Crystal structure of {\ETX} system. 
Conducting BEDT-TTF layers and insulating anion ones are stacked along the $b$ axis. 
Note that the crystal structure shown here is for {\RbZn} and {\CsZn}, and the long axis of the BEDT-TTF molecules in {\I} and {\aI} tilts slightly from the out-of-plane direction. 
(b) Molecular arrangement in conducting layers, in which BEDT-TTF molecules form an anisotropic triangular lattice. 
(c) Systematic variation in the charge-frustration parameters for the three {$\theta$}-phase materials.
(d) Configurations of current and voltage terminals for the present resistivity measurements.
}
\label{Fig1} 
\end{figure}

The organic materials, {\ETX} (BEDT-TTF denotes bis(ethylenedithio)tetrathiafulvalene), have conducting layers, in which BEDT-TTFs form isosceles triangular lattices dubbed $\theta$-type and accommodate a hole per two molecules, and insulating X (anion) layers \cite{Mori1998} [Fig.~1(a)].
 A strong Coulomb repulsion, $V$, between the neighboring molecular sites generally stabilizes a charge order (CO) \cite{Seo2000} but does not on the isosceles triangular lattices \cite{Seo2006, Kaneko2006}. 
The ratio of two intersite Coulomb energies,  $V_{\rm{p}}/V_{\rm{c}}$, a measure of charge frustration [Fig.~1(b)], is systematically varied with X as 0.87, 0.91 and 0.95 for X= RbZn(SCN)$_4$, CsZn(SCN)$_4$ and I$_3$ (hereafter abbreviated as {\RbZn}, {\CsZn} and {\I}), respectively [Fig.~1(c)] \cite{Mori2003, Kondo2006}. 
The least frustrated {\RbZn} exhibits a first-order transition into a horizontal stripe CO from a charge liquid (CL) at 200 K when cooled slowly at a rate of less than $\sim$5 K/min; however, a faster cooling suppresses the CO transition and gives a smooth pathway from the CL to the CG \cite{Kagawa2013, Nad2007, Nogami2010}.
 {\CsZn} does not exhibit the transition to the CO even when cooled slowly, e.g. at 0.1 K/min, but instead shows a continuous change from a CL to the CG similarly to the rapidly cooled {\RbZn} \cite{Sato2014, Nad2008, Suzuki2005}. 
Interestingly, the most frustrated system,{\I}, falls into a Fermi liquid (FL) instead of CG at low temperatures, arguably ascribable to a frustration-driven quantum melting of CG \cite{Sato2019, Merino2005, Tamura1994, Dressel2011}. 
The FL crosses over at high temperatures to a kind of CL \cite{Takenaka2005} distinguished from the CL in {\RbZn} and {\CsZn} in resistivity fluctuations and X-ray diffuse scattering \cite{Sato2019}; the nature of CL differs whether it falls into a CG or a FL at low temperatures.

The CG shares the universal features of conventional classical glasses, but raises a significant issue connected to the quantum nature of electron. 
The resistivity of the CG is weakly temperature-dependent, even metallic in {\CsZn}, and several orders of magnitude smaller than that of the CO. 
Obviously, a na\"ive picture of the CG, a glassy freezing of electron dynamics that should lead to the localization of electrons, does not hold in the real CG. 
The puzzling transport behavior is sure to be a key to the nature of the glass formed by the quantum particles, electrons.

Here, we investigate in-plane and out-of-plane resistivities for {\RbZn}, {\CsZn} and {\I} to characterize the fluidity of the CG in view of anisotropy, expecting that unusual charge excitations, if any, manifest themselves in directional dependence of transport characteristics as revealed in strongly correlated electron systems \cite{Nakamura1993, Valla2002}. 
A conventional metal-to-CO transition system, {\aETI} (hereafter abbreviated as {\aI}) \cite{Tajima2006, Liu2016} was also investigated as a reference of the frustration-free system. 
We found that the charge transport in the CG state is anomalously confined to 2D layers.

\begin{figure*}
\includegraphics[width=18.3cm,clip]{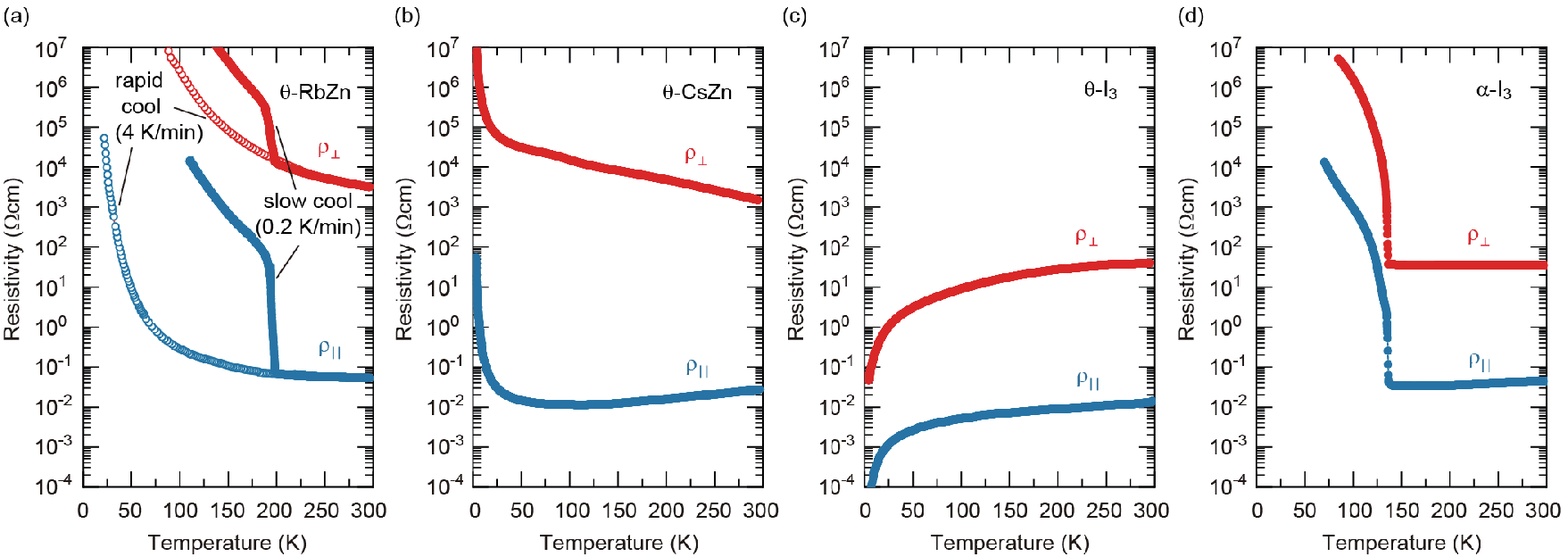}
\caption{(Color online) 
Temperature dependences of in-plane (blue) and out-of-plane (red) resistivities for (a) {\RbZn}, (b) {\CsZn}, (c) {\I} and (d) {\aI}. 
Open and closed circles in (a) represent the resistivities measured with rapidly (4 K/min) and slowly (0.2 K/min) cooled, respectively.
}
\label{Fig2} 
\end{figure*}

Single crystals of {\RbZn}, {\CsZn}, {\I} and {\aI} were synthesized by the galvanostatic anodic oxidation of BEDT-TTF, as described in the literatures \cite{Mori1998, Kajita1987}. 
Both the in-plane and out-of-plane resistivities were measured by the four-terminal method with the electrode configurations shown in Fig.~1(d); 
the in-plane resistivity was measured with the current contacts attached on the entire sides of a crystal to ensure an uniform current flow in the crystal, whereas the out-of-plane resistivity could be precisely measured with the conventional contacts of electrodes [Fig.~1(d)] thanks to a huge anisotropy of resistivity.

\begin{figure*}
\includegraphics[width=18.3cm,clip]{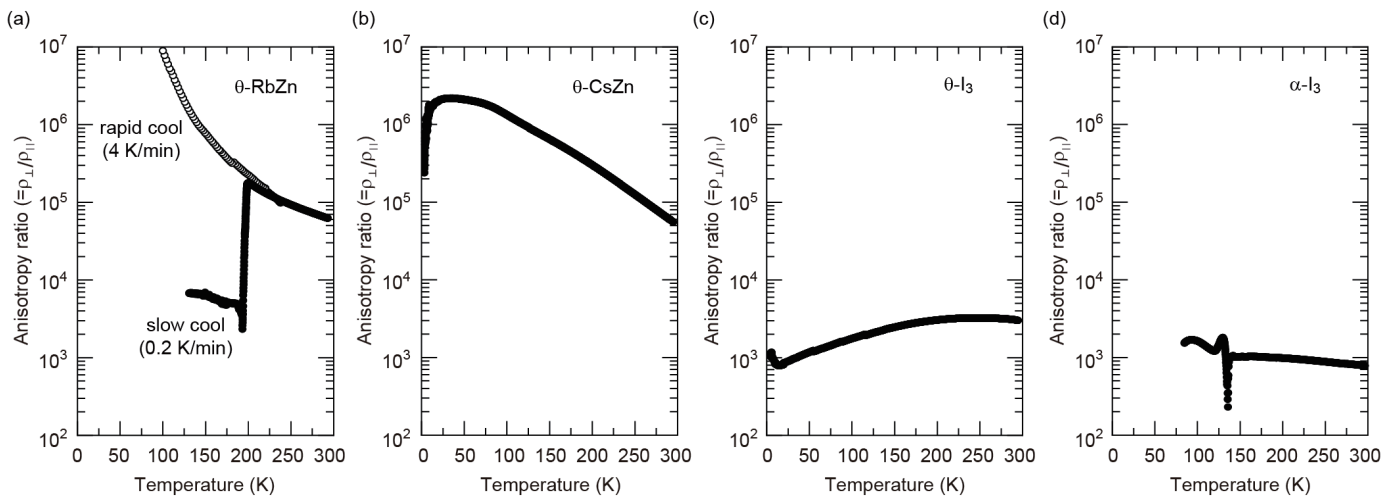}
\caption{(Color online) 
Temperature dependences of anisotropy ratios, {\ani}, for (a) {\RbZn}, (b) {\CsZn}, (c) {\I} and (d) {\aI}. 
Each profiles are deduced from the results in Fig.~2.
Open and closed circles in (a) represent the resistivities measured with rapidly (4 K/min) and slowly (0.2 K/min) cooled, respectively.
}
\label{Fig3}
\end{figure*}

Figures 2 shows the temperature dependence of out-of-plane resistivity, {\rhoout}, and in-plane resistivity, {\rhoin}, for the four systems. 
Abrupt increases in {\rhoout} and {\rhoin} at 200 K for {\RbZn} in a slowly cooled condition (0.2 K/min) indicate a transition from CL to CO, which is suppressed by rapid cooling (4 K/min) and gives way to a smooth change from CL to CG [Fig.~2(a)] similarly to the behavior observed in {\CsZn} irrespectively of the cooling rate [Fig.~2(b)]. 
The nominal CL to CG transition temperature, $T_{\rm{g}}$, is 160-170  K for {\RbZn} \cite{Kagawa2013} and 100 K  for {\CsZn} \cite{Sato2014} as the onsets of the ergodicity breaking; 
note that the $T_{\rm{g}}$ has no thermodynamic sense here but defines a phenomenological temperature at which the dynamics slows down to the laboratory time scale, e.g. $10^2$ s.
{\I} is metallic in the entire temperature range [Fig.~2(c)], and {\aI} shows a clear first-ordered CO transition both in {\rhoout} and {\rhoin} at 135 K [Fig.~2(d)]. 
A prominent feature of {\rhoout} and {\rhoin} of the CG states in {\RbZn} and {\CsZn} is a huge anisotropy of resistivity. 
The anisotropy ratio, {\ani}, is $(5-6)\times10^4$ for both systems in the CL state at room temperature, pointing to a highly 2D nature of the electronic state [Figs.~3(a) and (b)].
Remarkably, as temperature is lowered, {\ani} steeply increases up to the range of $10^6-10^7$ commonly for {\RbZn} and {\CsZn} [Figs.~3(a) and (b)].

The extraordinarily large values of {\ani} in CG and its anomalous temperature evolution over from CL to CG are highlighted with reference to the behavior of the metallic {\I, the CO state in {\RbZn} and the non-frustrated {\aI}. 
In {\I}, {\ani} is $3\times10^3$ at room temperature and gradually decreases with temperature to below $1\times10^3$ in the FL regime at low temperatures [Fig.~3(c)], where {\rhoin} varies in proportion to squared temperature \cite{Sato2019}. 
In the CO state of {\RbZn}, {\ani} falls into the range of $10^3$ [Fig.~3(b)]. 
Note that {\ani} in {\aI} is comparable to those of {\I} and the CO state of {\RbZn}, and remarkably nearly invariant across the metal-to-CO transition [Fig.~3(d)].
Thus, the electron transport strongly confined to the 2D layers is specific to the CG. 
We note that, in {\CsZn}, {\ani} turns to decrease at the lowest temperatures [Fig.~3(b)], coinciding with steep increases in both {\rhoout} and {\rhoin} [Fig.~2(b)]. 
This is likely related to the growing CO seeds in size ($\sim$60 \AA ~at 5 K) and fraction as signified by the X-ray diffuse scattering of (0, $k$, 1/2) at low temperatures \cite{Sato2014, Ito2008}.

The noticeable temperature dependence of {\ani} means that {\rhoout} and {\rhoin} vary differently with temperature. 
{\RbZn} is nonmetallic in both of {\rhoout} and {\rhoin}, so their behaviors are examined with the activation plots in Fig.~4. 
In the CL-CG state, the temperature dependence of {\rhoout}  is well described by an Arrhenius function with an activation energy, $\Delta$, of $\sim$1000 K in the form of {\rhoout}$\propto$$e^{\Delta /T}$, whereas {\rhoin} is characterized by $\Delta$$\sim$300 K at the low-temperature part (Fig.~4), indicating the strong suppression of the out-of-plane charge transport. 
We note that, in the CO state, {\rhoin} and {\rhoout} show a similar activation behavior with $\Delta$$\sim$1300 K and 1400 K, respectively, signifying no 2D confinement (Inset in Fig.~4).
The contrasting temperature dependence of {\rhoin} and {\rhoout} in the CL-CG state is particularly marked in {\CsZn}, where {\rhoin} behaves metallic whereas {\rhoout} is non-metallic in temperatures of 100-300 K; the highly conductive in-plane charge transport is prohibited in the out-of-plane direction.

The CL and CG in {\RbZn} exhibit short-ranged CO domains with a wave vector of {\boldmath $q_1$}[$\sim$($\pm$1/3, $k$, $\pm$1/4)] ($k$ denotes negligible coherence between the BEDT-TTF layers) \cite{Kagawa2013, Watanabe2004}. 
Theoretically, the long-ranged CO with the {\boldmath $q_1$} is stabilized by a strong charge frustration on triangular lattices and retains a metallic state because the periodicity is incommensurate for the 1/4-filled band \cite{Kaneko2006, Watanabe2006}. 
In reality, the {\boldmath $q_1$} modulation is not long-ranged but short-ranged {\boldmath $q_1$} domains are formed, which simultaneously cause glass-forming behavior and itinerancy of electrons. 
Note that the charge domains have no correlation between the adjacent layers \cite{Watanabe2003}, which should be partly responsible for the extremely 2D nature of the charge transport.
The contrasting temperature dependences of {\rhoin} and {\rhoout}, metallic v.s. nonmetallic, as observed in {\CsZn} may signal unconventional natures of low-energy charge excitations underlying their 2D confinement. 
In the Fermi liquid theory, which postulates quasiparticles as elementary excitations, the temperature dependence of out-of-plane conductivity should generally follow that of in-plane \cite{Valla2002, Ioffe1998}.
Therefore, the qualitative discrepancy between the temperature profiles of {\rhoin} and {\rhoout} suspects the conventional quasiparticle picture of the CL-CG state. 
The possible breakdown of the quasiparticle picture is also consistent with the fact that the metallic {\rhoin} exceeds the value of $\sim$1~m$\Omega$ cm deduced from $k_{F}l{\sim}1$ with $l$ the lattice spacing, the so-called Mott-Ioffe-Regel limit for the quasiparticle transport \cite{Takenaka2005}. 
The 2D confinement of charge transport is reminiscent of the similar behavior of the underdoped high-$T_{\rm{c}}$ copper oxides, in which charge and spin excitations are separated in the 2D conducting layers and hardly hop between the layers with the separated degrees of freedom retained. 
It is intriguing if separation or fractionalization of electron degrees of freedom or some collective excitations unable to directly hop to adjacent layers are associated with the 2D confined charge excitations; further insights await theoretical investigations.

\begin{figure}
\includegraphics[width=8.6cm]{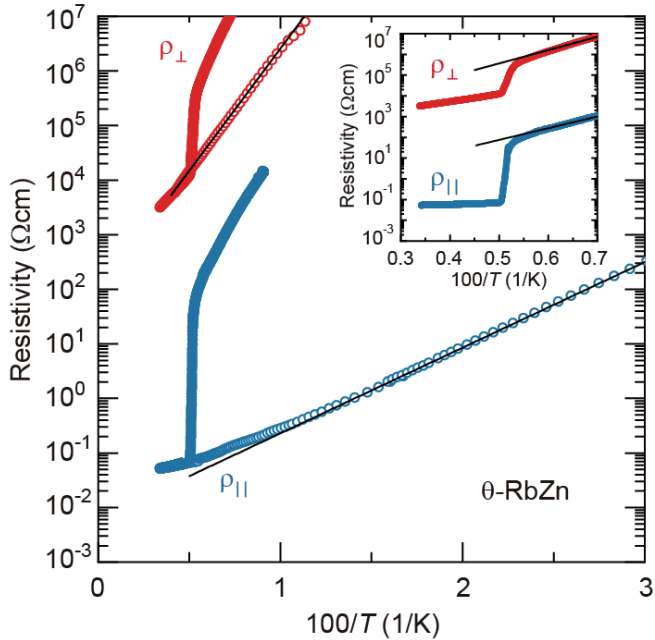}
\caption{(Color online) 
The in-plane (blue) and out-of-plane (red) resistivities for {\RbZn} against inverse of temperature. 
Open and closed circles represent the data acquired in the rapidly (4 K/min) and slowly (0.2 K/min) cooled conditions, respectively. 
The inset shows an enlarged figure around CL state. 
The black lines are resultant fitting curves by Arrhenius-type function.
}
\label{Fig4} 
\end{figure}

Apparently, the in-plane itinerancy of electrons in CG contradicts with the classical dynamics or freezing in glass; how do they reconcile with each other? 
As indicated by NMR spectra in {\RbZn} and {\CsZn} \cite{Sato2017, Chiba2008, Miyagawa2000}, the charge of the molecular site in the CG and CL states is distributed continuously in magnitude, indicating that the inhomogeneous entity responsible for the classical glass dynamics is the interaction-induced emergent \lq\lq charge density” instead of individual electrons. 
Specifically to the present systems, three-fold charge density modulations reconcile with itinerant electrons \cite{Kaneko2006, Watanabe2006, Hotta2006} and the itineracy is likely retained even when the three-fold modulation is short-ranged and takes on glassy feature due to the charge frustration.
The classical glass dynamics of the charge density should be reflected on the behavior of the itinerant fluid through their coupling. 
It is theoretically suggested that itinerant fermions with long-range interactions can form a glass state, dubbed a quantum charge glass \cite{Muller2012}. 
There is a general belief in the physics of soft mater that the hierarchical structure underlies the properties of conventional glass systems \cite{Palmer1984, Binder1986}.
We suggest that the electronic counterpart of the conventional glass adds the conceptually novel \lq\lq quantum-classical energetic hierarchy” to the physics of glass. 
The present results reveal highly unconventional low-energy excitations, which are strongly 2D confined and possibly distinct from the conventional quasi-particles, in the novel glass systems retaining quantum nature.

\begin{acknowledgments}
This work was supported in part by the Japan Society for the Promotion of Science 
(JSPS) KAKENHI (Grant Nos.~25220709, 17K05532, 18H05225 and 19H01846).
T.S. was supported as a JSPS Research Fellow (No. 14J07870).
\end{acknowledgments}

\end{document}